\documentclass[reprint,notitlepage,twocolumn,
superscriptaddress,aps,longbibliography,floatfix]{revtex4-1}
\usepackage[utf8]{inputenc}
\usepackage{braket}
\usepackage{amsmath}
\usepackage{amsthm}
\usepackage{mathrsfs}
\usepackage{mathtools}
\usepackage{amssymb}
\usepackage{dsfont}
\usepackage{braket}
\usepackage{bm}
\DeclareMathOperator{\tr}{tr}
\usepackage{color}
\pdfoutput=1

\usepackage[utf8]{inputenc}
\usepackage[english]{babel}
\usepackage[T1]{fontenc}
\usepackage{amsmath}
\usepackage{hyperref}
\usepackage[ruled,vlined]{algorithm2e}
\usepackage{enumitem}
\usepackage{amsthm}
\usepackage{amssymb}
\usepackage{graphicx}
\usepackage{titlesec}
\renewcommand{\thesection}{\arabic{section}}
\titleformat{\section}{\normalfont\bfseries\center}{Supplementary Note \thesection}{1em}{}

\usepackage{algorithm2e}

\begin{document}
\title{Quantum Similarity Testing with Convolutional Neural Networks}
\author{Ya-Dong Wu}
\affiliation{QICI Quantum Information and Computation Initiative, Department of Computer Science,
The University of Hong Kong, Pokfulam Road, Hong Kong}
\author{Yan Zhu}
\email{yzhu2@cs.hku.hk}
\thanks{Ya-Dong Wu and Yan Zhu contribute equally}
\affiliation{QICI Quantum Information and Computation Initiative, Department of Computer Science,
The University of Hong Kong, Pokfulam Road, Hong Kong}
\author{Ge Bai}
\affiliation{Centre for Quantum Technologies, National University of Singapore, Block S15, 3 Science Drive 2, 117543, Singapore}
\author{Yuexuan Wang}
\affiliation{AI Technology Lab, Department of Computer Science,
The University of Hong Kong, Pokfulam Road, Hong Kong}
\affiliation{ College of Computer Science and Technology,
Zhejiang University, Zhejiang Province, China}
\author{Giulio Chiribella}
\email{giulio@cs.hku.hk}
\affiliation{QICI Quantum Information and Computation Initiative, Department of Computer Science,
The University of Hong Kong, Pokfulam Road, Hong Kong}
\affiliation{Department of Computer Science, Parks Road, Oxford, OX1 3QD, United Kingdom}
\affiliation{Perimeter Institute for Theoretical Physics, Waterloo, Ontario N2L 2Y5, Canada}

\begin{abstract}

The task of testing whether two uncharacterized quantum devices behave in the same way is  crucial for benchmarking near-term quantum computers and  quantum simulators, but  has so far remained open for continuous-variable quantum systems. In this Letter, we develop a machine learning algorithm for comparing unknown continuous variable  states using limited  and noisy data.  The algorithm works on non-Gaussian quantum states for which similarity testing could not be achieved with previous techniques.   Our approach is based on a convolutional neural network that    assesses the similarity of quantum states based on a lower-dimensional state representation built  from  measurement data. The network can be trained offline with classically simulated data  from a fiducial set of states sharing structural similarities with the states to be tested,  or with  experimental data generated by measurements on the fiducial states, or with a combination of simulated and experimental data.  We test the performance of the model  on noisy cat states and states generated by arbitrary selective number-dependent phase gates.
  Our network  can also be applied to the problem of comparing continuous variable states across different experimental platforms, with different sets of achievable measurements, and to the problem of experimentally testing whether two states are equivalent up to  Gaussian unitary transformations.  
\end{abstract}
 
\maketitle

{\em Introduction.} Comparing  unknown quantum states based on  experimental data~\cite{elben2020,flammia2020,carrasco2021,daiwei2022} is crucial for  benchmarking   quantum simulations and near-term  quantum computers~\cite{eisert2020}.  
A natural approach in this context is to choose a trusted device as a reference standard, and to compare other   devices to it.  For example, the trusted  device could be built and maintained by a quantum computing company,  while the other devices could be owned by users  in distant laboratories. One way to  compare two unknown quantum devices is to   estimate their  overlap~\cite{buhrman2001,cincio2018,chabaud2018,fanizza2020,elben2020,guerini2021,anshu2022}, which is also useful  for tasks like  quantum state discrimination and classification~\cite{helstrom1969,holevo2011,sasaki2001,dunjko2018}.
Recently, Elben {\em et al.}\ proposed an approach named cross-platform verification~\cite{elben2020}, which uses  only local Pauli measurements to  experimentally  estimate the  overlap between two  multiqubit states.
 This approach has been recently demonstrated on quantum systems with more than ten qubits~\cite{zhu2022cross}.

An alternative approach to characterize quantum states from  measurement data is provided by deep neural networks~\cite{torlai2018np,Torlai2018prl,carrasquilla2019,torlai2019,tiunov2020,Ahmed2021PRL,smith2021,schmale2022,zhu2022,fedotova2022}, { which can work with a  smaller amount of data  when the states under consideration belong to state families  with sufficient structure, such as the family of ground states of the Ising model with different values of the couplings and of the magnetic field. } Recently, neural networks for  testing the similarity of  quantum states  have been developed for the tasks of  state discrimination and classification~\cite{magesan2015,carrasquilla2017,gao2018,cimini2020,Wetzel2020,zhang2021}. The existing methods, however,  generally assume  that the state of the system is identifiable by a finite set of predetermined labels.  For the purpose of comparison of uncharacterized quantum states, this assumption is too restrictive, as it prevents the  application  to continuous families of quantum states, such as the families of coherent states and cat states in quantum optics.

In this Letter, we develop a convolutional neural network  {  for testing the similarity of quantum states drawn from a  continuously-parametrized state family.  For sufficiently regular families, our network manages to tell different states apart using  noisy and incomplete measurement data,} without requiring randomization over an exponentially large set of measurements,  or correlations between  measurements performed on different states. The network  is trained with data  from a fiducial set of quantum states sharing structural similarities with the states to be compared. After training, the network embeds the measurement data into a low-dimensional feature space in a way that reflects the  similarity of quantum states. { This low-dimensional state representation is then used by the network to decide whether the two given states are the same or not.}  Our approach  is inspired by a classical technique  for the recognition  of human faces in blurred and incomplete images~\cite{schroff2015}, a task that shares similarities with the task of comparing continuous-variable quantum states from finite-statistics approximations of their Wigner function.  

  We test the performance of our network on noisy cat states, whose  preparation could not be  efficiently verified with previous techniques~\cite{Aol15,wu2021,elben2020}.    After an offline training on simulated data, the trained network is tested with both  simulated data and actual  experimental data.
    The  results indicate a  high success rate in disinguishing pairs of quantum states   using  incomplete measurement data that   are insufficient for a 
reliable state estimation with  maximum likelihood estimation~\cite{lvovsky2009} or generative adversarial networks~\cite{Ahmed2021PRL}.     { On the other hand, we observe that state families of larger effective size or larger complexity require a larger amount of data in order for the network to have  a satisfactory performance. 
 Our approach can also be used to verify   quantum states across different experimental platforms, having access to different types of  measurements in each platform.}
 In addition, it can also be used to test for the equivalence of quantum states up to a given set of  unitary operations, such as the set of Gaussian unitaries.  { Finally, this approach can be extended to families of discrete variable systems, such as the family of ground states of the  Ising model.}

{\em Framework.}
Two experimenters,  Alice and Bob, own two quantum devices  producing copies of two unknown  quantum states $\rho$ and $\sigma$,  respectively.  Alice and Bob want to determine whether their devices prepare the same quantum state, that is, whether $\rho =\sigma$. 
To this purpose, they  can only  perform  a  limited set of quantum measurements, possibly different for Alice and Bob.  In the following, we denote by $\mathcal M_A$  ($\mathcal M_B$) the set of measurements accessible to Alice (Bob). Each measurement $\bm M  \in  \mathcal M_A$  ($\bm M  \in  \mathcal M_B$) corresponds to a  positive operator-valued measure (POVM), that is, a set of positive operators  ${\bm M}:=\left  ( M_j\right)_{j=1}^k$ acting on the system's Hilbert space  and satisfying the normalization condition $\sum_{j=1}^k M_j =\mathds{1}$.
 To compare their states, Alice and Bob pick two subsets of measurements ${\cal S}_A  \subset \mathcal M_A$ and ${\cal S}_B  \subset \mathcal M_B$, respectively.  The measurements can be chosen independently and randomly, but there is no need for Alice and Bob to sample them from the uniform distribution, or from any specific probability distribution.      In general, the sets of performed measurements $\mathcal S_A$ and $\mathcal S_B$  need not  be informationally complete.

 By performing a   measurement $\bm{M} \in \mathcal S_A$  ($\bm{M} \in \mathcal S_B$) on  multiple  copies of $\rho$  ($\sigma$), Alice (Bob) obtains a vector of experimental frequencies $\bm{d}$. This classical data is then sent to a verifier, Charlie, whose task is  to decide whether $\rho$ and $\sigma$ are same state.   For each measurement    $\bm{M}  \in  \mathcal S_A$  ($\mathcal S_B$), Alice (Bob) provides Charlie with a pair $(\bm{m}, \bm{d})$, where  $\bm{m}$  is a parametrization of the measurement  $\bm{M}$.
 Here, the parametrization $\bm{m}$ could be either a full description of the POVM  $\bm{M}$, or a lower-dimensional parametrization valid only for measurements in $\mathcal M_A$ ($\mathcal M_B$).  

\begin{figure}
    \centering
    \includegraphics[width=0.45\textwidth]{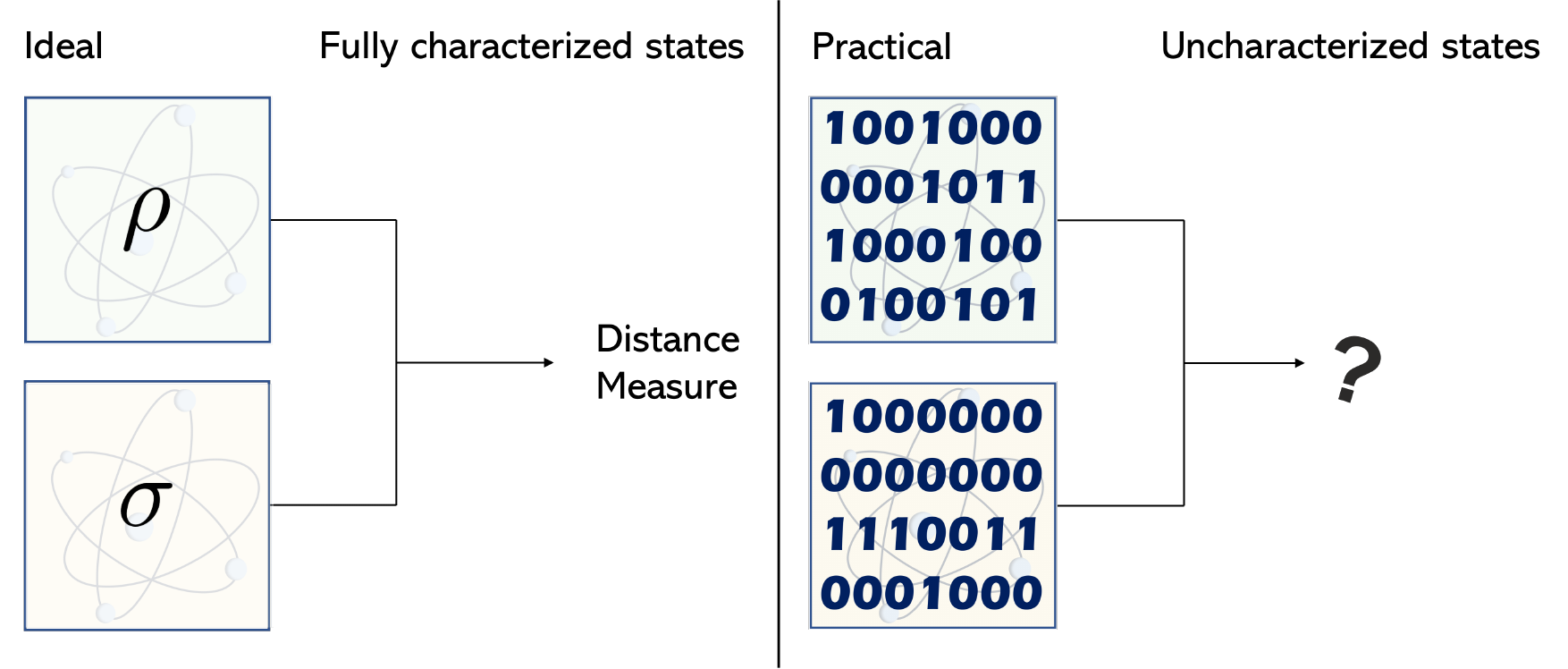}
    \caption{    Assessing the similarity of characterized vs uncharacterized quantum states.  For fully characterized quantum states (left), the similarity can be assessed by computing  their fidelity or any other distance measure.   Instead, for uncharacterized quantum states, only measurement data obtained from samples of the states  are available (right).  The task is then to determine whether  two given data sets have been generated from the same state or not. In general, the two data sets can refer to different sets of measurements,  and the latter can be noisy and  incomplete. }
    \label{fig:similarityLearning}
\end{figure}

Here we introduce a deep neural network  that determines whether two unknown states are same or not, using limited  and noisy data.
We call our network StateNet, in analogy to FaceNet~\cite{schroff2015}, a popular neural network for the identification of human faces.  
StateNet uses a convolutional neural network~\cite{lecun2015} to produce a low-dimensional representation of quantum states. For the training, we choose a set of fiducial states and  provide the corresponding measurement data to the network. { The training data can be  generated by computer simulation,  or by actual experiments, or by a combination of these two methods. Note that the training data need not  be produced afresh; instead, one can use existing data from past simulations or past experiments. } 
For each fiducial state $\tau$,   
its measurement data set is fed into a deep neural network to produce a low-dimensional representation, given by a vector $\bm{r}\in\mathbb R^n$. { The dimension  $n$ is a parameter of the network,  and its  choice is discussed in the Supplemental Material~\footnote{See Supplemental Material [url] for implementation and training details of our neural network model, which includes 
Refs.~\cite{aggarwal2018neural,johansson2012qutip,chollet2015keras,van2008visualizing}}. Note that in general $n$ can be much smaller than the dimension of the Hilbert spaces containing the fiducial states.}

In the training phase, we optimize the parameters of the convolutional neural network with respect to a loss function, called triplet loss~\cite{schroff2015}. After the training is concluded, the  network maps  measurement data to vectors that reflect the similarity of quantum states, as illustrated in Supplementary Note 3.
Quantum states are then compared by evaluating the Euclidean distance between the corresponding  vectors.  To decide whether two vectors correspond to the the same quantum state, the network uses a  threshold value that balances between the false rejection rate and the false acceptance rate over a new set of unseen measurement data obtained from the fiducial states.  
The  details of StateNet and its training are presented in Supplementary Note 1.

{\em Testing the similarity of  continuous variable  states.}
  A continuous-variable quantum state $\rho$ is characterized by its Wigner function~\cite{PhysRevLett.78.2547,PhysRevLett.89.200402} $W_\rho(\alpha):=\frac{2}{\pi}\tr(\rho D(\alpha)(-1)^{\hat{n}}D(-\alpha))$, where $D(\alpha)  :=  \exp[  \alpha a^\dag -  \bar\alpha\, a]$ is a displacement operator, $\hat{n}=\hat{a}^\dagger\hat{a}$ is the photon number operator, and $\hat{a}^\dagger$ and $\hat{a}$ are the bosonic creation operator and annihilation operator respectively.   An estimate of the value of $W_\rho(\alpha)$ at any phase-space point $\alpha$ can be achieved  {\em e.g.} by measuring displaced parity operator $D(\alpha)(-1)^{\hat{n}}D(-\alpha)$, which is widely used for the characterization of  quantum states in circuit quantum electrodynamics~\cite{vlastakis2013,kudra2022,heeres2015,sivak2022}.

Suppose that Alice (Bob) can  estimate the Wigner function  at a finite number of points, chosen at random from a square grid over the phase space.   The set of all  points on the grid corresponds to the set of achievable measurements   $\mathcal M_A  =  \mathcal M_B  =: \mathcal M$.
Alice (Bob) randomly chooses a subset of points %vertices
in the grid to perform measurements of the Wigner function, which is a subset $\mathcal S_A$  ($\mathcal S_B$)   of $\mathcal M$. In general, the points chosen by Alice and Bob  need not be the same.  
After a finite set of measurement runs, Alice (Bob) obtains a two-dimensional data image, where some pixels are missing and
the value at each of the existing pixels is an estimate of the Wigner function at the associated phase-space point.  As a result of the finite statistics, the image will generally be blurred.

\begin{figure}
    \centering
    \includegraphics[width=0.35\textwidth]{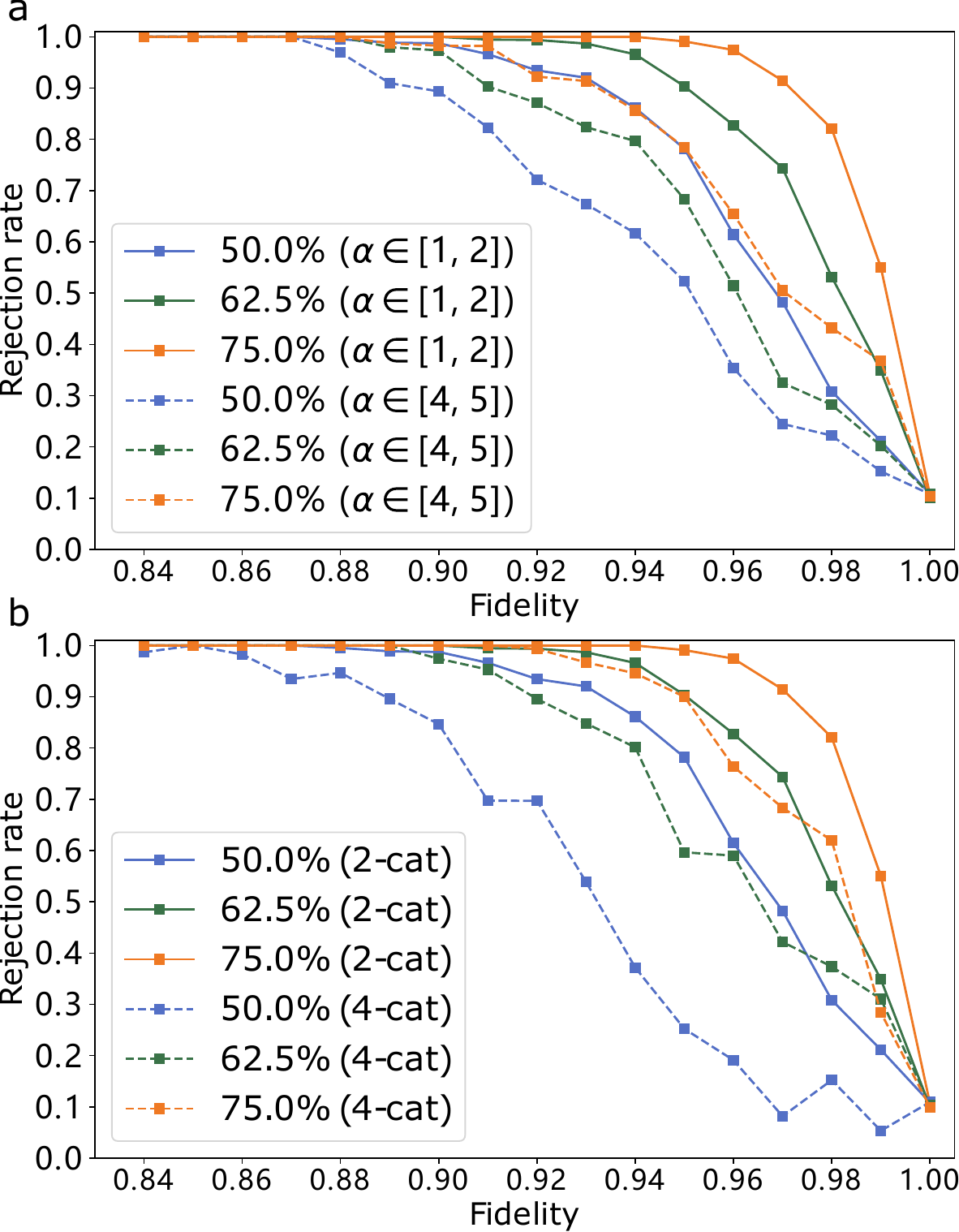}
    \caption{      Rejection rate for cat states as  a function of the fidelity.   Three different scenarios are considered: $50\%$, $62.5\%$ and $75\%$
of $32 \times 32$ pixels are randomly selected as input to the neural
network respectively. For all the scenarios, each pixel is an estimate of
the Wigner function obtained by sampling the true outcome
probability distribution $300$ times. 
 In Fig.(a), solid lines are for cat states with $\alpha \in[1,2]$ and dashed lines are for cat states with $\alpha\in [4,5]$. In Fig.(b),  solid lines are for four-component cat states and dashed lines are for two-component cat states.}
    \label{fig:rejectionRate}
\end{figure}

We test our method on  cat states \cite{yurke1986,mirrahimi2014} 
    $(\ket{\alpha}+\ket{-\alpha})/{\sqrt{2(1+\exp(-2|\alpha|^2))}}$,  
where $\ket{\alpha}   :=  D(\alpha)  |0\rangle$ is a coherent state.   
We train StateNet using simulated measurement data from  ideal cat states  as well as  noisy cat states with a fixed amount of thermal noise. After the training is concluded, we test the performance of StateNet in distinguishing between pairs of noisy cat states degraded by  photon loss. For each pair of noisy cat states $\rho$ and $\sigma$, we take $\rho$ to play the role of the reference state, and make it close to its noiseless counterpart $\rho_{\text{ideal}}$, with fidelity 99\%).  On the other hand, we regard  $\sigma$  as the untrusted state that needs to be verified, and allow it to be generally noisier, allowing the fidelity with the trusted state $\rho$ to range between $84\%$ and $100\%$.
  To evaluate the performance of the network, we plot the rejection rate, namely the probability that  the two states are judged to be different, as a function of their fidelity.   The resulting plot is shown in Fig.~\ref{fig:rejectionRate}.

 Fig.~\ref{fig:rejectionRate}(a) shows the performance of StateNet for   noisy cat states with amplitudes $\alpha\in [1, 2]$ and $\alpha\in [4, 5]$. 
The numerical results indicate that, when the amplitude is increased without increasing the amount of measurement data,  the prediction accuracy decreases. 
 We also test how the performance of our neural network is affected by the states' complexity, as measured by their  nonclassicality~\cite{kenfack2004}. 
 To this purpose,  we  consider cat-like states that are superposition of four coherent states instead of two. 
Fig.~\ref{fig:rejectionRate}(b) demonstrates that the success rate of StateNet decreases as state complexity increases. 

 \begin{figure}
    \centering
    \includegraphics[width=0.45\textwidth]{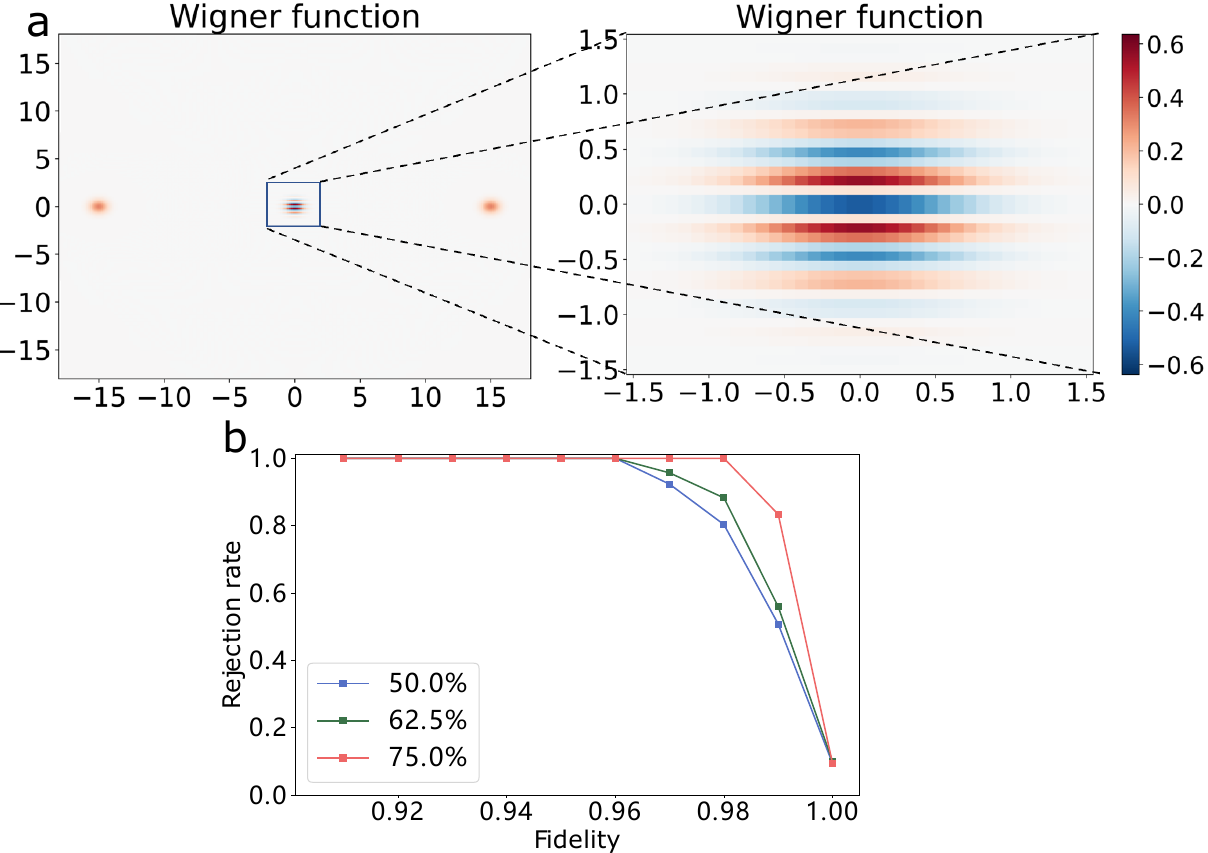}
    \caption{ 
    (a) Wigner function of a cat state with $\alpha=16$, along with a inset figure of the Wigner function on a $36\times 36$ fine grid within the region $[-1.5, 1.5]\times [-1.5, 1.5]$.
  (b) rejection rates against quantum fidelity when StateNet only utilizes $50\%$, $62.5\%$, $75\%$ of the measurement data of the Wigner function on this $36\times 36$ fine grid.}
    \label{fig:largeCat}
\end{figure}

{
To test the ability of our network to cope with high-dimensional quantum states, we perform numerical experiments on  noisy cat states with high amplitudes. 
Figure~\ref{fig:largeCat}(b) illustrates the performances of StateNet for the comparison of two noisy cat states with amplitudes $\alpha\in [15,16]$ (corresponding to an average number of photons between $225$ and $256$) using displaced parity measurement data on a $36\times 36$ grid within $[-1.5, 1.5]\times [-1.5, 1.5]$ in phase space. Despite the limited amount of measurement data, StateNet achieves a relatively high success rate in this high amplitude scenario. }

In addition to cat states, we also consider the family of states  of the form
\begin{equation}\label{eq:snapstate}
\ket{\phi_{{\bm \theta}, \alpha}}:=\text{SNAP}(\bm{\theta})\ket{\alpha},
\end{equation}
where $\text{SNAP}(\bm{\theta}):=\sum_{n} \exp(\text{i}\theta_n )\ket{n}\bra{n}$ is a selective number-dependent arbitrary phase gate~\cite{heeres2015,fosel2020}.
 In this example, we take $\theta_0 = \theta_1  =  \pi$ and $\theta_n  =  0$ for every $n\ge 2$.
For the training, we used computer  simulated measurement data. For the testing, instead, we used actual experimental  data obtained from a noisy implementation~\cite{Ahmed2021PRL,kudra2022} using  a resonator coupled to a superconducting transmon qubit.  For the test data sets, we randomly choose   $5\%$ of $81\times 81=6561$  experimental data pixels corresponding to the same quantum state (shown in Fig.4(b) in Ref.~\cite{Ahmed2021PRL}), and we   check whether StateNet correctly attributes  two such  data sets to  same quantum state. %{ [I rephrased the previous sentence, Plase check if it is correct. ]}
The  acceptance rate, averaged over multiple trails, is $98\%$. 
%StateNet can also be adapted to determining whether two quantum devices implement the same quantum dynamics, as investigated in Supplementary Note 4.  { [given that none of the referees seems to care about this result, how about removing it from this paper, and making a separate paper on the similarity testing for quantum dynamics? If we just put the result here in the Supplemental Material, I doubt that anyone will notice it. ]}

{\em  Verification of Equivalence up to Gaussian Unitary Operations.}
 A variant of StateNet can be used to decide whether two quantum states are the same up to a unitary transformation in a given set. 
 This functionality  can be used to make our comparison  method   robust to unknown systematic unitary errors taking place on individual quantum devices. A common example of such
errors are the errors introduced by poorly calibrated displacement pulses, phase rotations and squeezing~\cite{weedbrook2012}. 

\begin{figure}
    \centering
    \includegraphics[width=0.48\textwidth]{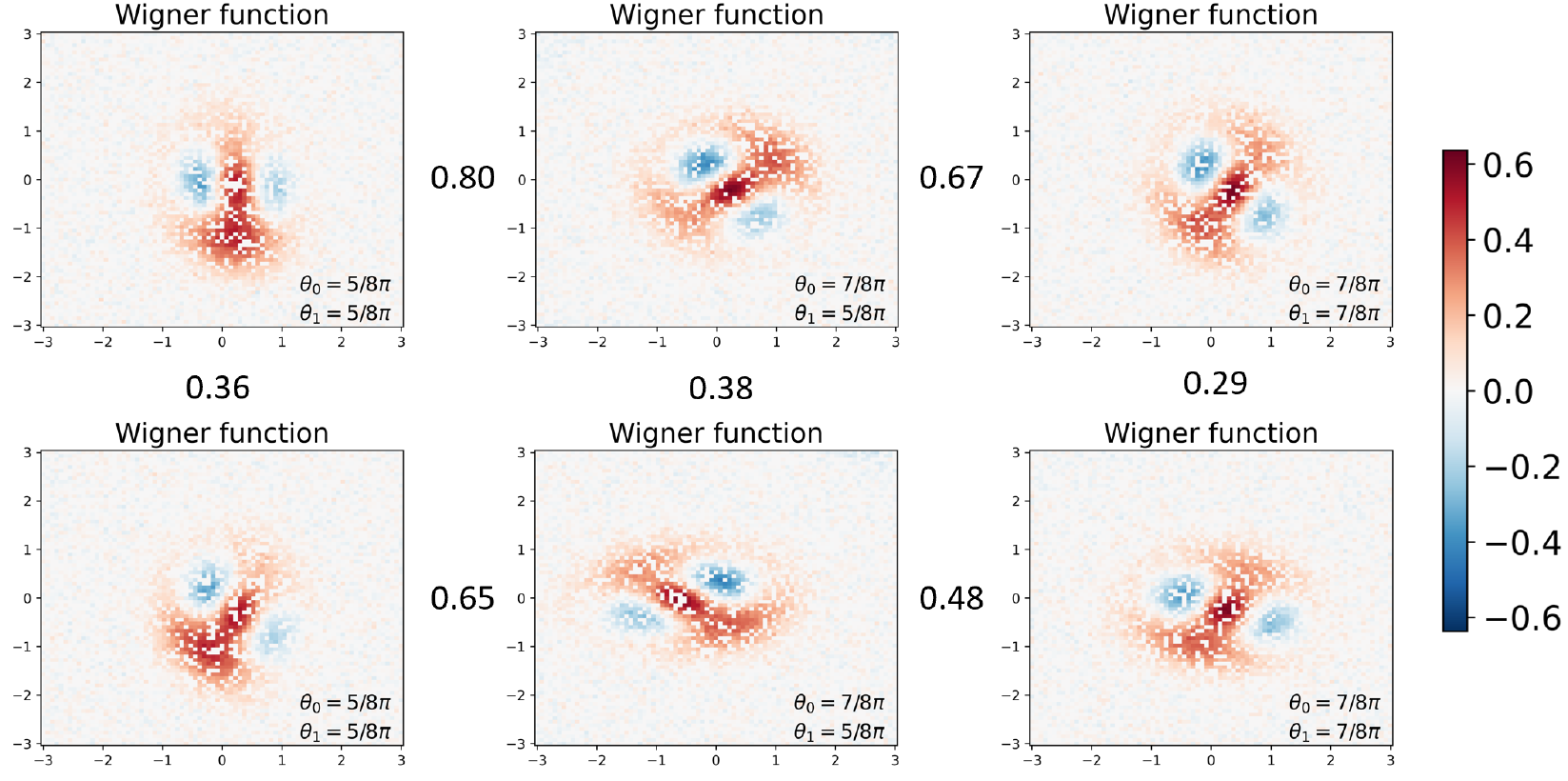}
    \caption{ 
      Verification of equivalence up to Gaussian unitary transformations. Each column contains a pair of data images of the same quantum state with different affine transformations. The three different columns correspond to three quantum states (\ref{eq:snapstate}) with different values of $\theta_0$ and $\theta_1$ ($\theta_n=0$ for $n\ge 2$). The number between each pair of data images is the Euclidean distance between their state representations produced by StateNet. Each  data image contains 4900 pixels randomly chosen from $81\times 81=6561$ pixels, and the value at each pixel is an estimate of the Wigner function obtained by sampling the outcome probability distribution for $500$ times.}
    \label{fig:affine}
\end{figure}

In the Wigner function representation, the combination of displacements, rotations, and squeezing corresponds to an affine transformation in  phase space.  We use StateNet for testing whether two data images generated from states of the form~(\ref{eq:snapstate}) are equivalent  up to an affine transformations. 
Fig.~\ref{fig:affine} shows examples of image data for equivalent as well as inequivalent quantum states.
 By balancing both the false rejection rate and the false acceptance rate, we obtain a distance threshold $0.4$, which makes our model accept all pairs of equivalent states, and reject the inequivalent ones. 
 More numerical results are shown in Supplementary Note 4.

\begin{figure}
    \centering
    \includegraphics[width=0.4\textwidth]{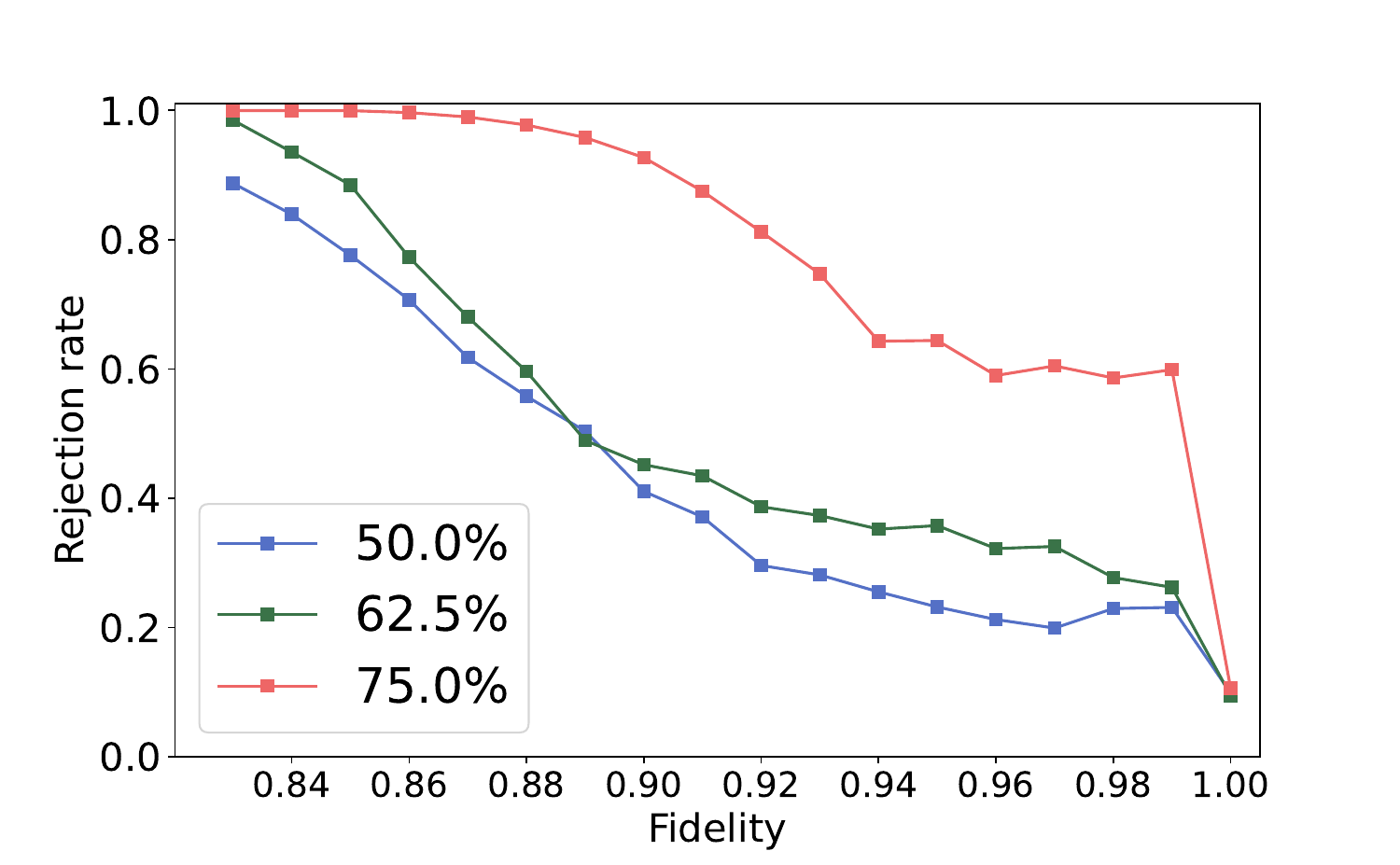}
    \caption{ Rejection rate as a function of quantum fidelity for comparing cat states with homodyne measurement data and displaced parity measurement data. For displaced parity measurements, $50\%$, $62.5\%$ and $75\%$ of $32\times 32$ pixels are randomly selected as input to the
neural network respectively, where each pixel is
an empirical average of $300$ samples. For homodyne measurements, $32$ quadrature phases are randomly selected from $[0, \pi)$ and each quadrature phase corresponds to a frequency distribution of $1440$ samples.}
    \label{fig:crossPlatform}
\end{figure}

{
{\em Similarity testing with two different  sets of achievable measurements. }
Quantum information  protocols have been implemented on a variety of experimental platforms,  each involving different sets of feasible  measurements. 
For instance,  homodyne measurements are commonly used for photonic systems~\cite{lvovsky2009}, while displaced parity measurements are preferable in cavity quantum electrodynamics~\cite{heeres2015,sivak2022}. 
Here we demonstrate that StateNet test the similarity of quantum states realized on two different experimental platforms, with two different sets of accessible measurements.

We consider a scenario where Alice performs displaced parity measurements on  state $\rho$, while Bob performs homodyne measurements on  state $\sigma$. 
Given measurement data from these two essentially different types of measurements, Charlie aims to determine whether $\rho$ equals to $\sigma$. To achieve this objective, we jointly train two neural networks, so that their measurement data from these two different types of measurements are mapped into a single  representation space. 
Fig.~\ref{fig:crossPlatform} illustrates the rejection rates against quantum fidelity between two noisy cat states.}

{
{\em Similarity testing for multiqubit states.}
StateNet can also be adapted to the problem of testing the similarity of multiqubit states, such as the ground states of Ising model. 
We test its performance on  pairs of $10$-, $20$-or $50$-qubit states $\rho$ and $\sigma$, where $\rho$ represents an ideal ferromagnetic Ising ground state, and  $\sigma$ an untrusted Ising ground state with poor calibration of the coupling parameters.
Alice (Bob) selects  a subset of two-qubit  nearest-neighbor Pauli measurements, and the  measurement statistics is then  input into StateNet. The results of our  numerical experiments, provided in Supplementary Note 5, show that  our approach can correctly identify whether two datasets are from the same $\rho$ or different states $\rho$ and $\sigma$ with a probability of over $90\%$ for 10 qubits. However, for $50$ qubits, the success probability drops to $80\%$ if the amount of measurement data is kept fixed.  }

{\em Conclusions.}  We introduced a model of neural network for  testing the similarity of quantum states from limited noisy data. 
%Our model has two important benefits:   {\em  (i)} applicability to states from   continuous state families, 
%and {\em (ii)} no need of correlated randomized measurements.
Our work opens up the application of  metric learning techniques~\cite{schultz2003} to the  characterization of quantum systems produced by noisy intermediate-scale quantum devices~\cite{preskill2018}. 
Furthermore, it sheds light on how  machines  can discover physical notions, such as the similarity of two quantum states,  without any  hard-coded information about the corresponding  physical theories ~\cite{iten2020,flam2022,krenn2022}.

{\em Acknowledgement.}
This work was supported by funding from the Hong Kong Research Grant Council through grants no.\ 17300918 and no.\ 17307520, through the Senior Research Fellowship Scheme SRFS2021-7S02, the Croucher Foundation, and by  the John Templeton Foundation through grant  62312, The Quantum Information Structure of Spacetime (qiss.fr). 
YXW acknowledges funding from the National Natural Science Foundation of China through grants no.\ 61872318. Research at the Perimeter Institute is supported by the Government of Canada through the Department of Innovation, Science and Economic Development Canada and by the Province of Ontario through the Ministry of Research, Innovation and Science. The opinions expressed in this publication are those of the authors and do not necessarily reflect the views of the John Templeton Foundation.

\bibliography{refs.bib}

\begin{widetext}
\section*{\uppercase{ Supplementary Information}}
\section{Implementation details of StateNet}
\subsection{Structure of StateNet}

As shown in Supplementary Fig.~\ref{fig:snet_struc}, our proposed StateNet for learning quantum state similarity consists of a representation network $f_{\bm{\xi}}$ and a $L_2$ normalization layer. 

\begin{figure}[h]
    \centering
    \includegraphics[width=0.45\textwidth]{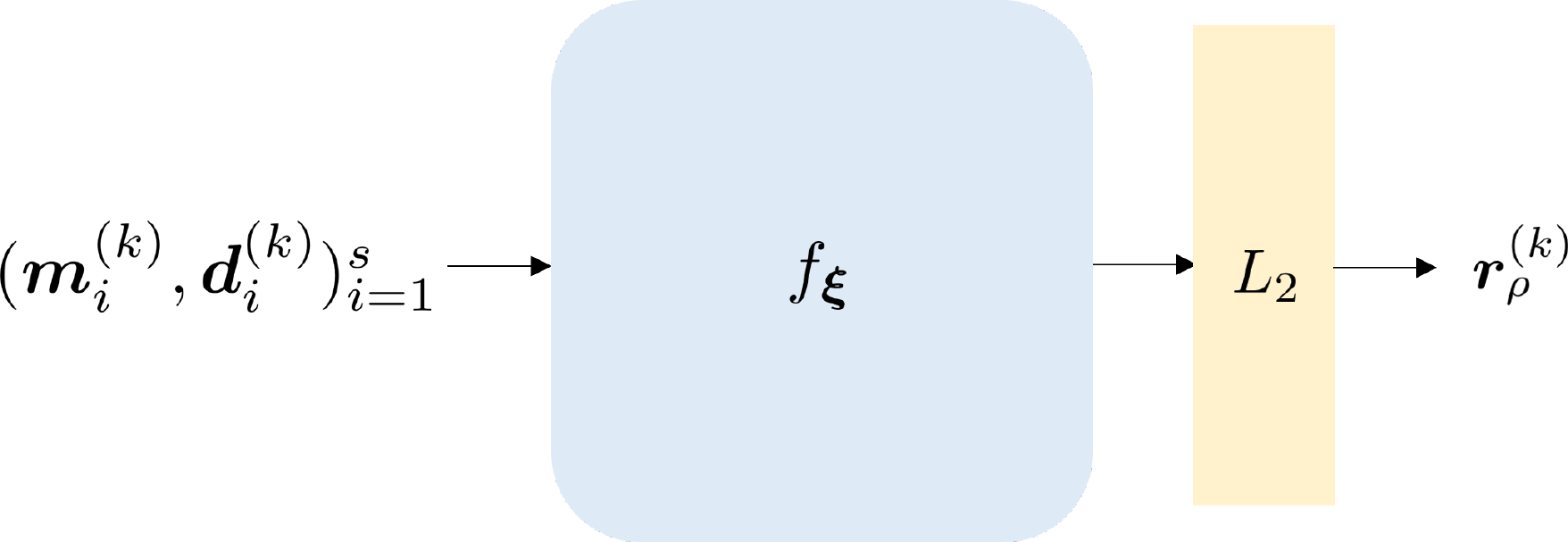}
    \caption{Structure of StateNet.}
    \label{fig:snet_struc}
\end{figure}

The representation network $f_{\bm{\xi}}$ is composed of multiple convolutional layers, a max pooling layer, a dropout layer and a last full-connected layer  \cite{aggarwal2018neural} and we depict its structure in Supplementary Fig.~\ref{fig:rep_struc}. We denote all trainable parameters in the representation network as $\bm{\xi}$.  The input of the representation network is a set of $s$ measurement results of a quantum state and we denote them as $\left(\bm{m}_j^{(k)}, \bm{d}_j^{(k)}\right)_{j=1}^s$, where $\bm{m}_j$ is parameterization of the measurement and $\bm{d}_j$ is a vector of experimental measurement outcome frequencies on the state. A $L_2$ normalization layer following the representation network is used for rescaling the output of the representation network to have Euclidean norm $1$. 

\begin{figure}[h]
    \centering
    \includegraphics[width=0.35\textwidth]{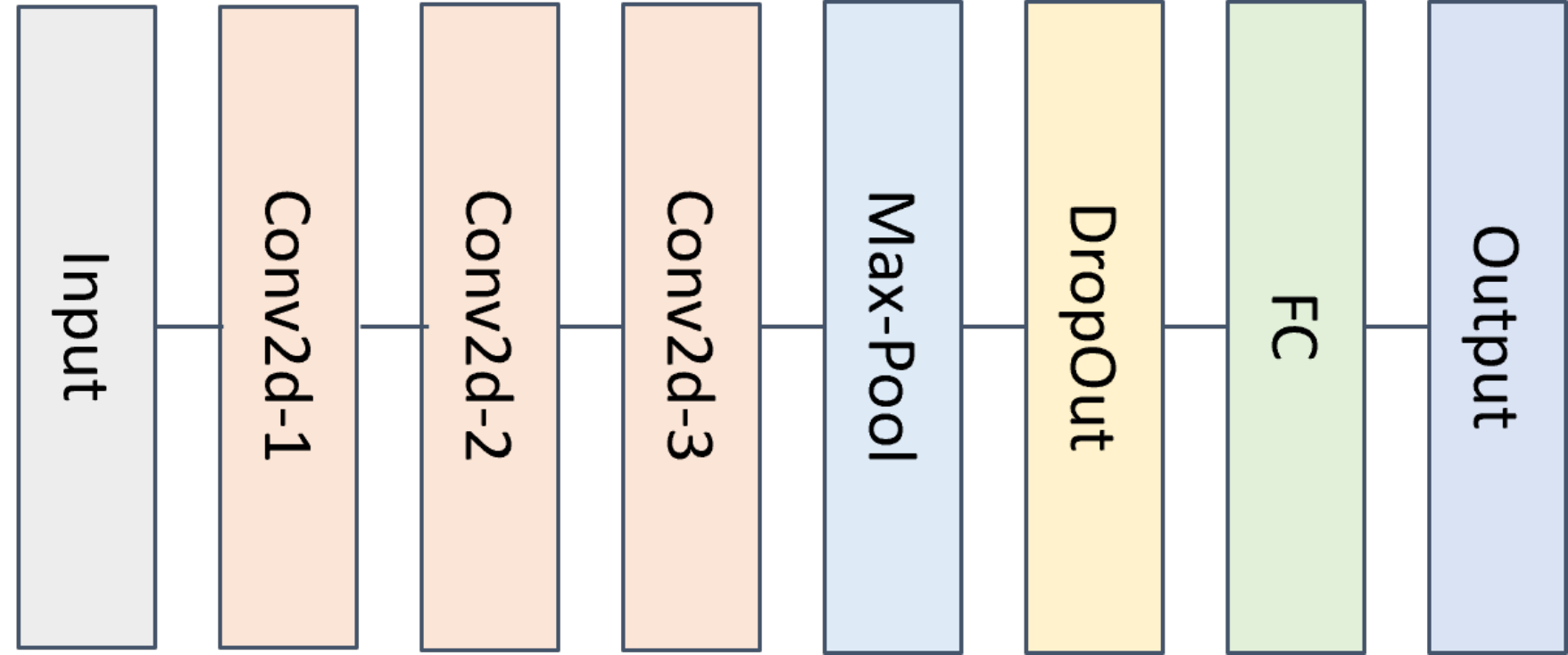}
    \caption{Structure of the representation network.}
    \label{fig:rep_struc}
\end{figure}

\subsection{Training Details}

\paragraph{Loss function.} In the training, we adopt a loss function called Triplet Loss used in FaceNet. The Triplet Loss minimizes the distance between an anchor and a positive, both of which belong to the same quantum state, and maximizes the distance between the anchor and a negative which belongs to different quantum states.
In each round of optimization, we  choose a representation $\bm{r}_\tau^{(l)}$ as a reference with a  fixed  state $\tau$ and $l$. Then,  we compare the Euclidean distances $||\bm{r}_\tau^{(l)}-\bm{r}_\tau^{(i)}||_2$   of state representations associated to different measurements on the same state $\tau$,  with the Euclidean distances 
$||\bm{r}_\tau^{(l)}-\bm{r}_\xi^{(m)}||_2$ of state representations associated to different states $\tau$ and $\xi$, for every $1\le i \le K$,  $i\not =  l$,  $1\le m\le K$, and for a randomly chosen fiducial state $\xi\neq \tau$.
Optimizing triplet loss aims to minimize $||\bm{r}_\tau^{(l)}-\bm{r}_\tau^{(i)}||_2$ for multiple rounds of experiments with respect to same fiducial state $\tau$, and simultaneously maximize $||\bm{r}_\tau^{(l)}-\bm{r}_\xi^{(m)}||_2$ for each pair of different fiducial states $\tau\neq \xi$. 
The detailed steps of training is presented in Algorithm  \ref{algo:training_statenet}.

\begin{algorithm}
\caption{Training of StateNet for learning quantum state similarity.}\label{algo:training_statenet}
% \DontPrintSemicolon
\KwData{number of states in training set $N$, state measurement results $\{\{\chi_{\rho_{i}}^{(k)}\}_{k=1}^{K}\}_{i=1}^{N}$, maximum number of epochs $E$, learning rate $\delta$}

Initialize parameters $\bm{\xi}$ and randomly, $e = 0$\;
\While{$e<E$}{
    Calculate the state representation $r_{\rho_i}^{(k)}$ for each state measurement result $\chi_{\rho_{i}}^{(k)}$\;
    $\mathcal{L} = 0$\;
    \For{$i_1=1$ \KwTo $N$}{
        Randomly select $K$ state representations $r_{\rho_j}^{(k)}$ from all state representations, where $j\neq i_1$, and denote them as $\{r_{\rho_{neg}}^{(k)}\}_{k=1}^K$ \;
        \For{$i_2=1$ \KwTo $K$}{
            Select $r_{\rho_{i_1}}^{(i_2)}$ as a reference state representation\;
            Calculate the Euclidean distance $\text{dist}_k:=|| r_{\rho_{i_1}}^{(i_2)}- r_{\rho_{i_1}}^{(k)}||_2$ between $r_{\rho_{i_1}}^{(i_2)}$ and each $r_{\rho_{i_1}}^{(k)}$, where $k \in \{1,2,\cdots,K\}$\;
            Calculate the Euclidean distance $||r_{\rho_{i_1}}^{(i_2)}-r_{\rho_{neg}}^{(k)}||_2$ between $r_{\rho_{i_1}}^{(i_2)}$ and each element in $\{r_{\rho_{neg}}^{(k)}\}_{k=1}^K$ and find the minimum  $\text{dist}_{\text{neg}}:=\min_{k}||r_{\rho_{i_1}}^{(i_2)}-r_{\rho_{neg}}^{(k)}||_2$\;
            $\mathcal{L} = \mathcal{L} + \sum_{k=1}^{K}\text{dist}_k -K\cdot \text{dist}_{\text{neg}}$ \;
        }
    }
    Update $\bm{\xi}$  as $\bm{\xi} = \bm{\xi}-\delta \nabla_{\bm{\xi}} \mathcal{L}$\;
    $e = e + 1$ \; 
    } 
\end{algorithm}

\paragraph{Training set and validation set.} Both the training and the validation sets are simulation data in our experiments. We usually generate all of the simulated data and apportion the data into training and validation sets, with an $80$-$20$ split.  The model is first trained over the training set and the threshold is then chosen over the validation set by minimizing both the false rejection rate and the false acceptance rate.

\paragraph{Initialization and learning rate.} We randomly initialize the parameters of the StateNet in each experiment and set the initial learning rate as $0.01$. We decreases the learning rate as the number of iterations increases. 

\paragraph{Number of epochs and training time.} The maximum number of epochs $E$ is usually set as $500$ in the training. The training time depends on the specific dataset and tasks we consider but it is less than one hour in all of the experiments shown in this paper.

\paragraph{Dimension of state representations.} The dimension of the state representation space is set to be $n=32$ for all the numerical experiments. Changing the dimension of state representations affects the performance of our data-driven approach.  When we reduce the dimension of state representations from $32$ to $8$, we find clear reduction of success rates in distinguishing two noisy states only when quantum fidelity is above $0.9$ and below one, as shown in Fig.~\ref{fig:representationDimension}.

\begin{figure}
    \centering
    \includegraphics[width=0.45\textwidth]{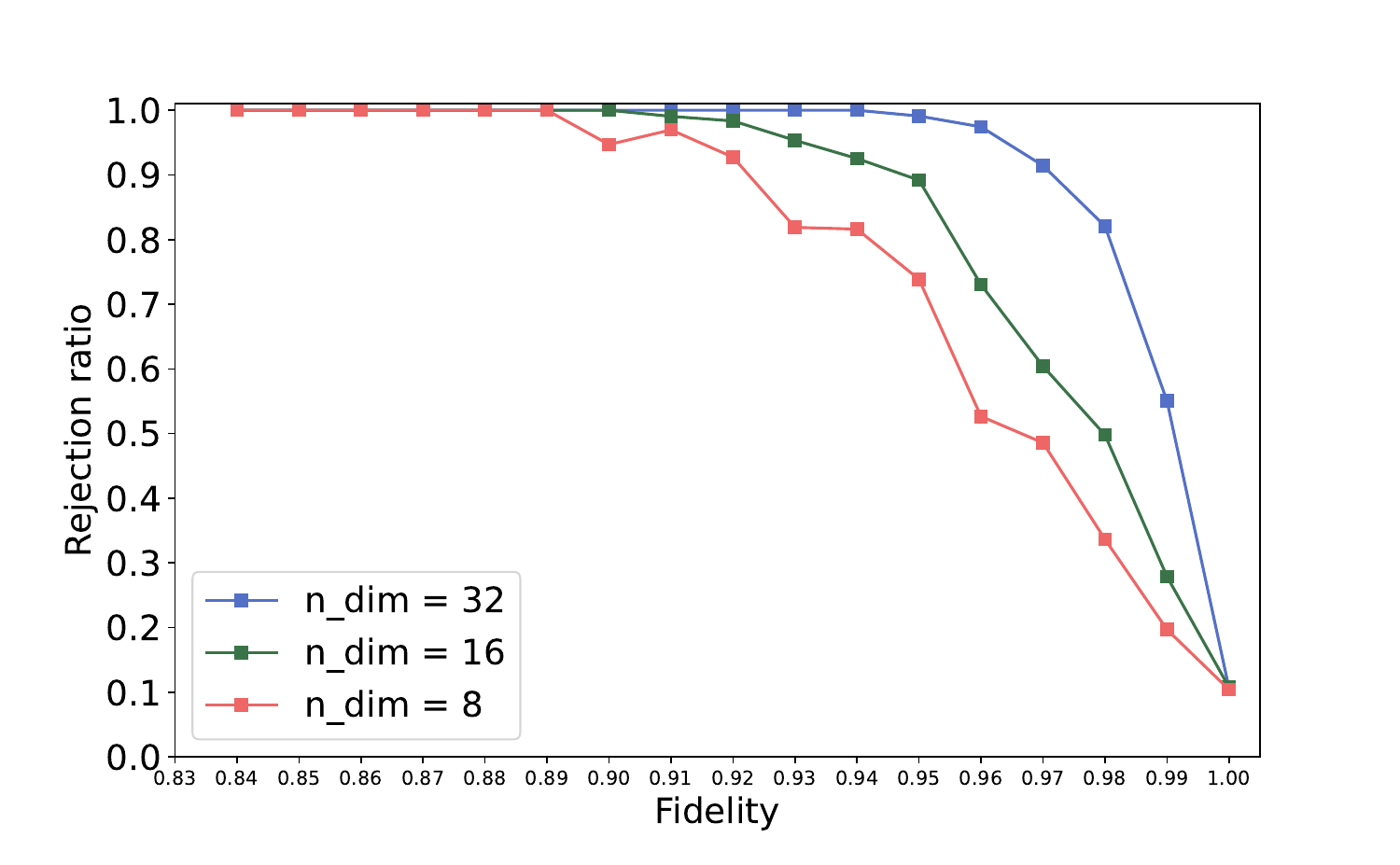}
    \caption{The rejection rates as functions of quantum fidelity between two noisy cat states with respect to different state representation dimensions $n=32$, $16$ and $8$.}
    \label{fig:representationDimension}
\end{figure}

\section{Dataset}
In this section, we will introduce the types and the generation method of the fiducial states used in the settings we consider.
\paragraph{Cross-Platform Verification of Quantum States} The training set is composed of the simulated measurement data of both ideal cat states in Eq. (1) with $\alpha \in \{1,1.1,\cdots, 2\}$ and noisy cat states with certain amount of thermal noises. The test set is composed of noisy cat states with photon loss errors. In order to test our model for noisy experimental data, we also use simulated measurement data of states in Eq. (2) with $\alpha \in \{\frac{\pi}{2},\frac{5\pi}{8},\cdots, \frac{3\pi}{2}\}$ to train our proposed model. All of these states are generated by qutip \cite{johansson2012qutip}. 
\paragraph{Cross-Platform Verification of Quantum Dynamics} In this experiment, fiducial states in the training set is composed of all the states at different time step $t_i$ when the states are initialized in coherent states with $\alpha \in \{1,1.1,\cdots,2\}$. All of these states are generated by qutip \cite{johansson2012qutip}. The test set consists of quantum states under the noiseless quantum evolution and the noisy quantum evolution.
\paragraph{Cross-Platform Verification of Equivalence up to Unitary
Operations} In this setting, our proposed StateNet is trained by the measurement data corresponding to the states in Eq. (2) with $\theta_0, \theta_1\in \{\pi/2, 3\pi/4, \cdots, 3\pi/2\}$ and tested by the  measurement data corresponding to the states with $\theta_0, \theta_1\in\{5\pi/8, 7\pi/8, \cdots, 11\pi/8\}$. The states are generated by qutip \cite{johansson2012qutip} and the affine transformations are simulated by tools provided in Keras \cite{chollet2015keras}.

\section{Visualization of State Representations}
It is also useful to visualize the state representations obtained by embedding different measurement data in feature space.  To this purpose, we feed the state representations of training states with $\alpha\in\{1, 1.1, \dots, 1.9\}$, into
a t-distributed stochastic neighbor embedding (t-SNE)
algorithm~\cite{van2008visualizing} to project the representation vectors into a two-dimensional plane, according to their similarities in feature space.
The results, provided in Fig.~\ref{fig:tsne}, show that measurement data associated to the same state effectively correspond to nearby vectors, while measurement data associated to different states typically correspond to distant vectors.  It is worth stressing that  
  StateNet  was not provided  with  any state parameter ({\em e.g.} it was not provided the value of $\alpha$), nor it was provided any information about the fidelity of  quantum states.  The notion of distance visualized in the figure was developed  by the network  as a way to minimize the loss function in the training phase. 
 %Autonomously the network learns a metric distance over the manifold of quantum states in Eq.~(\ref{eq:cat}) with $\alpha\in [1,2)$.

\begin{figure}
    \centering
    \includegraphics[width=0.35\textwidth]{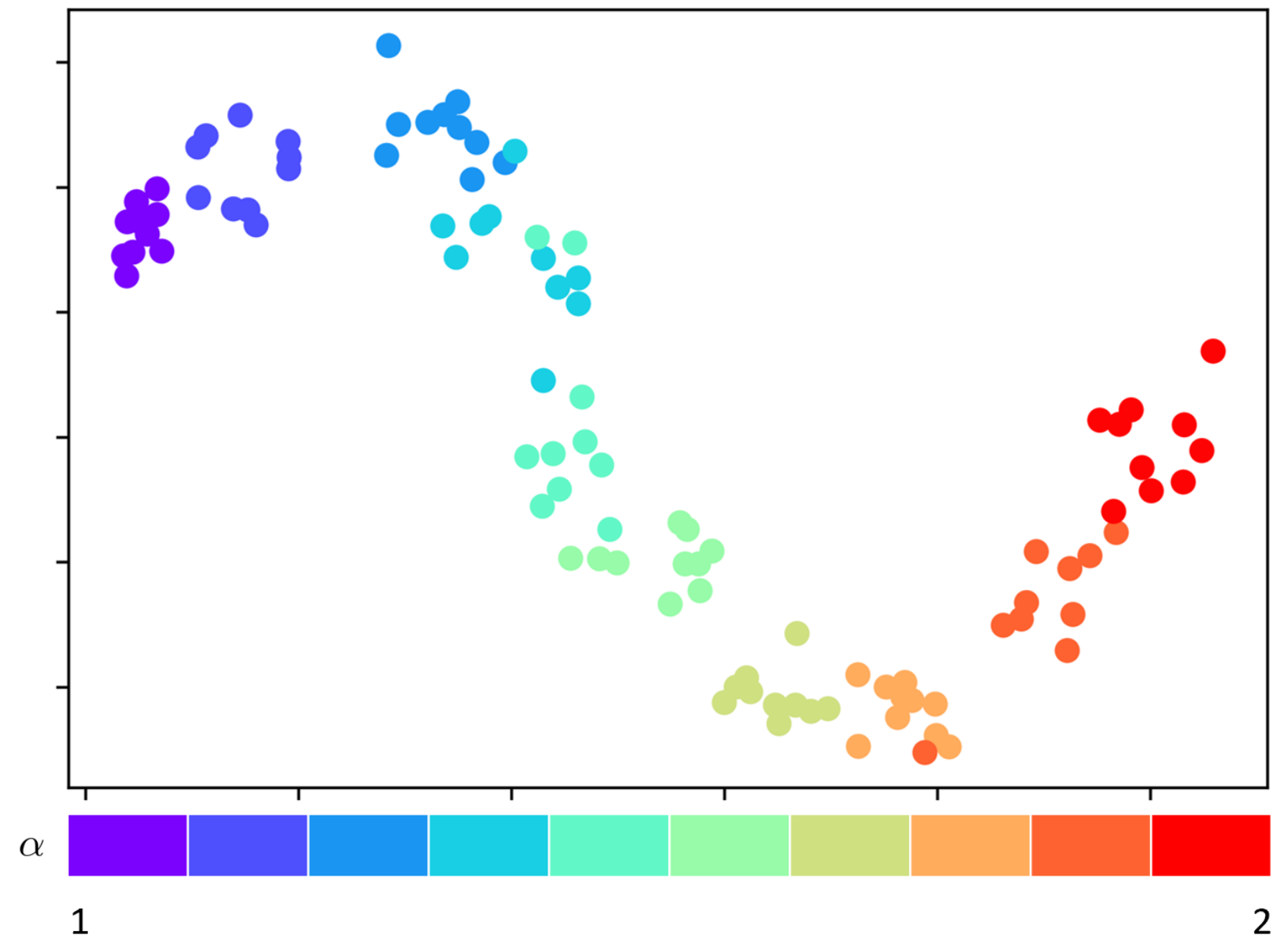}
    \caption{2D projections of state representations corresponding to cat states $\propto \ket{\alpha}+\ket{-\alpha}$ with $\alpha\in\{1, 1.1, \dots, 1.9\}$ obtained by t-SNE.  In the figure, points of the same color correspond to state representations associated to different measurements on the same quantum state. }
    \label{fig:tsne}
\end{figure}

\section{Verification of Equivalence up to Gaussian Unitary Operations}
In the Wigner function representation, the combination of displacements, rotations, and squeezing corresponds to the following affine transformation in  phase space 
\begin{align*}
    \begin{pmatrix}
    x \\
    p
    \end{pmatrix}
    \rightarrow 
    \begin{pmatrix}
    \zeta(x\cos \phi+ x \sin \phi+\Delta_x) \\
    1/\zeta(p\sin \phi+p \cos \phi+\Delta_p)
    \end{pmatrix},
\end{align*}
where $\phi$ is a rotation angle, $\Delta_x$ and $\Delta_p$ are shifts in position and momentum, respectively, and $\zeta$ is a squeezing parameter.  
%To simulate the distortion effect induced by systematic measurement errors, 
Each  data image, with respect to quantum states (1) in main text with $\theta_0, \theta_1\in [\pi/2, 3\pi/2]$ and $\theta_n=0$ for $n\ge 2$, undergoes an affine transformation with uniformly random parameters $\phi\in [0, \pi)$, $\Delta_x, \Delta_p \in (-1, 1)$ and $\zeta\in [5/6, 6/5]$. Then data images with same values of $\bm{\theta}$ but different affine transformations are labeled the same, while data images with different values of $\bm{\theta}$ are identified with different labels.

We train StateNet with data corresponding to $\theta_0, \theta_1\in \{\pi/2, 3\pi/4, \cdots, 3\pi/2\}$ and then test it over unseen data images corresponding to $\theta_0, \theta_1\in\{5\pi/8, 7\pi/8, \cdots, 11\pi/8\}$ to decide whether two data images correspond to the same quantum state up to certain distortion effect.
Numerical results show that both the average rate to reject different labels and the average rate to accept same labels is around $90\%$. 
%Fig.~\ref{fig:affine} shows examples of image data with both the same and different labels,  where the values between pairs of images are the Euclidean distances between the corresponding state representations produced by StateNet.   
% By balancing both the false rejection rate and the false acceptance rate over a validation set, we obtain a distance threshold $0.4$, which makes our model accept all pairs of states that are same up to an affine transformation, and reject those that belong to different classes. 

\section{Verification of Ising ground states}
In this section, we apply our method to verification of many-qubit states. We consider $10$-, $20$- and $50$-qubit ground states of ferromagnetic Ising model. The Hamiltonian of Ising model is
\begin{equation}\label{transIsing}
    H=-\left(\sum_{i=0}^{L-2} J_i\sigma_i^z\sigma_{i+1}^z+\sum_{j=0}^{L-1} \sigma_j^x\right),
\end{equation}
where coupling parameters $J_i>0$ correspond to ferromagnetic  interactions. 
Here the measurement set  $\cal M$ consists of all two-qubit Pauli measurements on  nearest-neighbor qubits.
Alice (Bob) can randomly
choose a subset of measurements from $\mathcal{M}$ and we denote it as $\mathcal{S}_A(\mathcal{S}_B)$. Note that $\mathcal{S}_A$ can be different from  $\mathcal{S}_B$. After a finite set of measurement
runs, Alice (Bob) will collect $|\mathcal{S}_A|(|\mathcal{S}_B|)$ measurement outcome probability distributions corresponding to $\mathcal{S}_A(\mathcal{S}_B)$. 

We train StateNet using simulated measurement data from  ideal ground states of the Ising model in Eq.~(\ref{transIsing}). After the training is concluded, we test the performance of StateNet in the  cross-verification of pairs of states $\rho$ and $\sigma$. For each pair, we randomly choose $J\in [0.3 ,0.4)$ and make $\rho$ an ideal Ising ground state with $J_i=J$, which plays the role of the reference state.  On the other hand, we regard  $\sigma$  as the untrusted state that needs to be verified, and make it degraded by poor calibration of coupling parameters, where $J_i$ becomes a Gaussian random variable with mean value $J$ and variance $0.1$.
In Supplementary Table~\ref{tab:Ising}, we show the acceptance rates for the same ideal ground states and the rejection rates for different states in the test. 

%In order to validate our proposed model, we generate $100$ different . For each generated state, we choose 10 different $\mathcal{S}_A(\mathcal{S}_B)$and collect their corresponding probability distributions. We expect our StateNet to distinguish different states from these probability distributions.  We further split these generated states into training set and test set with ratio 80:20. 

\begin{table}[h]
    \centering
    \begin{tabular}{c|c|c|c}\hline\hline
         Acceptance/rejection rates & $10$-qubit & $20$-qubit & $50$-qubit  \\ \hline
         $50\%$ of measurement data & 0.96/0.93 & 0.91/0.87 & 0.84/0.80 \\ \hline
         $62.5\%$ of measurement data & 1/1 & 0.95/0.93 & 0.89/0.89\\ \hline
         $75\%$ of measurement data & 1/1 & 0.96/0.93 & 0.92/0.90 \\ \hline\hline
    \end{tabular}
    \caption{The acceptance rates for same state and the rejection rates for different states with respect to $10$-, $20$-, and $50$-qubit ground states of Ising model when $\mathcal S_A$ and $\mathcal S_B$ consist of $50\%$, $62.5\%$ and $75\%$ of $\mathcal M$.}
    \label{tab:Ising}
\end{table}
\end{widetext}
\end{document}